\documentclass[12pt,preprint]{aastex}
\usepackage{underscore}
\usepackage{color}

\shorttitle{Polarimetry for 209P/LINEAR} 
\shortauthors{Kuroda et al.}

\begin{document}

\title{Optical and Near-Infrared Polarimetry for a Highly Dormant Comet 209P/LINEAR}

\author{Daisuke \textsc{Kuroda}}
\affil{Okayama Astrophysical Observatory, National Astronomical Observatory of Japan, Asakuchi, Okayama 719-0232, Japan}

\author{Masateru \textsc{Ishiguro}}
\affil{Department of Physics and Astronomy, Seoul National University, Gwanak, Seoul 151-742, Korea }

\author{Makoto \textsc{Watanabe}}
\affil{Department of Cosmosciences, Graduate School of Science, Hokkaido University, Kita-ku, Sapporo 060-0810, Japan}

\author{Hiroshi \textsc{Akitaya}\altaffilmark{\dag}}
\affil{Hiroshima Astrophysical Science Center, Hiroshima University, Higashihiroshima, Hiroshima 739-8526, Japan }

\author{Jun \textsc{Takahashi}}
\affil{ Nishi-Harima Astronomical Observatory, Center for Astronomy, University of Hyogo, Sayo, Hyogo 679-5313, Japan}

\author{Sunao \textsc{Hasegawa}}
\affil{Institute of Space and Astronautical Science (ISAS), Japan Aerospace Exploration Agency (JAXA), Sagamihara, Kanagawa 252-5210, Japan}

\author{Takahiro \textsc{Ui}, Yuka \textsc{Kanda}, Katsutoshi \textsc{Takaki}, Ryosuke \textsc{Itoh}}
\affil{Department of Physical Science, Hiroshima University, Higashihiroshima, Hiroshima 739-8526, Japan}

\author{Yuki \textsc{Moritani}\altaffilmark{\ddag}}
\affil{Kavli Institute for the Physics and Mathematics of the Universe (WPI), The University of Tokyo, 5-1-5, Kashiwanoha, Kashiwa, 277-8583, Japan}

\author{Masataka \textsc{Imai}, Shuhei \textsc{Goda}}
\affil{Department of Cosmosciences, Graduate School of Science, Hokkaido University, Kita-ku, Sapporo 060-0810, Japan}

\author{Yuhei \textsc{Takagi}, Kumiko \textsc{Morihana}, Satoshi \textsc{Honda}}
\affil{Nishi-Harima Astronomical Observatory, Center for Astronomy, University of Hyogo, Sayo, Hyogo 679-5313, Japan}

\author{Akira \textsc{Arai}\altaffilmark{\S}}
\affil{Koyama Astronomical Observatory, Kyoto Sangyo University, Motoyama, Kamigamo, Kita-ku, Kyoto 603-8555, Japan}

\author{Hidekazu \textsc{Hanayama}}
\affil{Ishigakijima Astronomical Observatory, National Astronomical Observatory of Japan, 1024-1 Arakawa, Ishigaki, Okinawa 907-0024, Japan}

\author{Takahiro \textsc{Nagayama}}
\affil{Graduate School of Science and Engineering, Kagoshima University, Kagoshima 890-0065, Japan}

\author{Daisaku \textsc{Nogami}}
\affil{Department of Astronomy, Graduate School of Science, Kyoto University, Kyoto 606-8502, Japan}

\author{Yuki \textsc{Sarugaku}}
\affil{Kiso Observatory, Institute of Astronomy, School of Science, The University of Tokyo, Kiso-gun, Nagano 397-0101, Japan}

\author{Katsuhiro \textsc{Murata}}
\affil{Department of Astrophysics, Nagoya University, Chikusa-ku, Nagoya 464-8602, Japan}

\author{Tomoki \textsc{Morokuma}}
\affil{Institute of Astronomy, Graduate School of Science, The University of Tokyo, Mitaka, Tokyo 181-0015, Japan}

\author{Yoshihiko \textsc{Saito}}
\affil{Department of Physics, Tokyo Institute of Technology, Meguro-ku, Tokyo 152-8551, Japan}

\author{Yumiko \textsc{Oasa}}
\affil{Faculty of Education, Saitama University, Sakura, Saitama 338-8570, Japan}

\author{Kazuhiro \textsc{Sekiguchi}, Jun-ichi \textsc{Watanabe}}
\affil{National Astronomical Observatory of Japan, Mitaka, Tokyo 181-8588, Japan}

%\altaffiltext{\dag}{Department of Cosmosciences, Graduate School of Science, Hokkaido University, Kita-ku, Sapporo 060-0810, Japan}
\altaffiltext{\dag}{Core of Research for the Energetic Universe, Hiroshima University,Higashi-Hiroshima, Hiroshima 739-8526, Japan}
\altaffiltext{\ddag}{Hiroshima Astrophysical Science Center, Hiroshima University, Higashihiroshima, Hiroshima 739-8526, Japan}
\altaffiltext{\S}{Nishi-Harima Astronomical Observatory, Center for Astronomy, University of Hyogo, Sayo, Hyogo 679-5313, Japan}

\begin{abstract}
We conducted an optical and near-infrared polarimetric observation of the highly dormant Jupiter-Family Comet, 209P/LINEAR.
Because of its low activity, we  were able to determine the linear polarization degrees of the coma dust particles and nucleus independently,
that is $P_n$=30.3$^{+1.3}_{-0.9}$\% at $\alpha$=92.2\arcdeg\ and $P_n$=31.0$^{+1.0}_{-0.7}$\% at $\alpha$=99.5\arcdeg\ for the nucleus, and $P_c$=28.8$^{+0.4}_{-0.4}$\% at $\alpha$=92.2\arcdeg\ and 29.6$^{+0.3}_{-0.3}$\% at $\alpha$=99.5\arcdeg\ for the coma.
We detected no significant variation in $P$ at the phase angle coverage of 92.2\arcdeg--99.5\arcdeg, which may imply that the obtained polarization degrees are nearly at maximum in the phase-polarization curves. 
By fitting with an empirical function, we obtained the maximum values of linear polarization degrees $P_\mathrm{max}$=30.8\% for the nucleus and $P_\mathrm{max}$=29.6\% for the dust coma. The $P_\mathrm{max}$ of the dust coma is consistent with those of dust-rich comets. The low geometric albedo of $P_v$=0.05 was derived from the slope--albedo relationship and was associated with high $P_\mathrm{max}$.
We examined $P_\mathrm{max}$--albedo relations between asteroids and 209P, and found that the so-called Umov law seems to be applicable on this cometary surface.
\end{abstract}


%%%%%%%%%%%%%%
%     INTRODUCTION     %
%%%%%%%%%%%%%%
\section{Introduction}
\label{sec:introduction}
% 300--400 words
The linear polarization of light scattered by airless solar system objects (i.e. comets and asteroids) is a useful tool for investigating the physical properties of their surfaces. 
The phase-polarization curves display a common behaviour, having a negative polarization branch at 0\arcdeg$<$$\alpha$$\lesssim$20\arcdeg, positive branch at $\alpha \gtrsim$20\arcdeg, and maximum polarization around $\alpha$=90--100\arcdeg, where $\alpha$ denotes the solar phase angle (i.e. Sun--object--observer's angle). 
The phase-polarization curves of asteroids, usually observed in the mainbelt at a low phase angle ($\alpha <$30\arcdeg), provide information about composition (i.e. taxonomic type), optical properties,  porosity of the surface regolith layers, and so on \citep[see e.g.][]{Dollfus1989,Shkuratov2002,GilHutton2014}.
In contrast, comets, which are usually enclosed in a dust coma plus gas contamination, have been observed with polarimeters at a wide range of phase angles, providing the composition, size and structure (fluffy or compact) of dust grains in comae \citep{Zubko2011,Kolokolova2007}.

Although several near-Earth asteroids (NEAs) and active comets were observed at large phase angles, little is known about the polarimetric properties of bare comet nuclei at large phase angles. 
Whenever comets are observed at large phase angles (i.e. $\alpha>$90\arcdeg), their nuclei are supposed to shrouded in thick cometary comae, because comets are located within 1 AU in the geometry where they are  heated up, creating outflow of dust particles and sublimating ice.
This paper attempted to obtain unique data of the linear polarization degree, $P$, for a bare cometary nucleus as well as dust particles of 209P/LINEAR (hereafter 209P). 
The comet is classified among the Jupiter-Family Comets (hereafter JFCs, the semimajor axis $a$=2.932 AU, eccentricity $e$=0.692, inclination $i$=19.4\arcdeg, and the Tisserand parameter with respect to Jupiter, $T_\mathrm{J}$=2.80) and known as the parent body of a meteor shower, the May Camelopardalids \citep{Jenniskens2014}. 
It closely encountered  the Earth in late May, 2014, providing us with the  opportunity to observe it at large phase angles.
\citet{Ishiguro2015} noticed that 209P nucleus was largely mantled by the surface dust layer, showing very weak activity when it closely approached Earth in 2014 April--May.

We conducted polarimetric observations of the comet at large phase angles in optical and near-infrared wavelengths using newly developed instruments as part of {\it OISTER} (an inter-university observation network in the optical and infrared wavelengths) activities. From the observed data, we subtracted the faint coma components to derive the nuclear $P_n$, and compare the results with those of previous studies.

%%%%%%%%%%%%%%%%%%%%%%%%%%%%%%%%
%     SECTION 2: OBSERVATION AND DATA REDUCTION    %
%%%%%%%%%%%%%%%%%%%%%%%%%%%%%%%%
\section{Observations and Reductions}

% 2.1. Observation
\subsection{Observations}
\label{subsec:observations}
Polarimetric observations of 209P were conducted for five nights in 2014 April--May using two telescopes: the Hiroshima Optical and Near-InfraRed camera (HONIR) on the 1.5-m Kanata telescope (hereafter Kanata) at the Higashi-Hiroshima Astronomical Observatory, Hiroshima, and the visible Multi-Spectral Imager (MSI) on the 1.6-m Pirka telescope (Pirka) at Hokkaido University's Nayoro Observatory in Hokkaido, Japan. 
In the imaging polarimetric mode, each  instrument employs a Wollaston prism beam-splitter and a rotatable half-wave plate modulator, which produces reliable data sets taking ordinary and extraordinary images simultaneously at four position angles $\theta$=0\arcdeg, 45\arcdeg, 22.5\arcdeg, and 67.5\arcdeg.
HONIR provides a simultaneous optical and near-infrared imaging polarimetry in a sequence of exposures. 
The detectors consist of a 2048 $\times$ 4096 pixel Hamamatsu Photonics fully depleted back-illuminated CCD with a pixel scale of 15 \micron\ for optical channel and a 2048 $\times$ 2048 pixel Raytheon VIRGO-2K HgCdTe array with a pixel scale of 20 \micron\ for infrared channel, respectively \citep{Akitaya2014}. MSI enables  optical polarimetric measurements using the EM-CCD camera (Hamamatsu Photonics C9100-13), which employs a back-thinned 512 $\times$ 512 pixel frame transfer CCD with a pixel scale of 16 \micron\ \citep{Watanabe2012}.
We observed 209P with HONIR/Kanata and MSI/Pirka at large phase angles, that is, 85.5--99.5\arcdeg\ (R$_\mathrm{C}$-band) and 85.5--89.9\arcdeg\ (J-band). We employed a non-sidereal (cometary motion) tracking with MSI/Pirka, and a sidereal tracking with HONIR/Kanata. The exposure times were chosen ranging from 45 seconds to 180 seconds depending on the detected fluxes and the apparent motion of 209P. Note that the apparent motion of the comet was fast in late May 2014 (5--7\arcsec\ min$^{-1}$ with respect to the sidereal tracking) so that we chose short exposure times to reduce the risk of tracking errors of these telescopes.  Details of those observations are summarized in Table \ref{tab:t1}. 

% 2.2 Data Reduction
\subsection{Data Reductions}
\label{subsec:reductions}
\begin{figure}
 \epsscale{1.0}
   \plotone{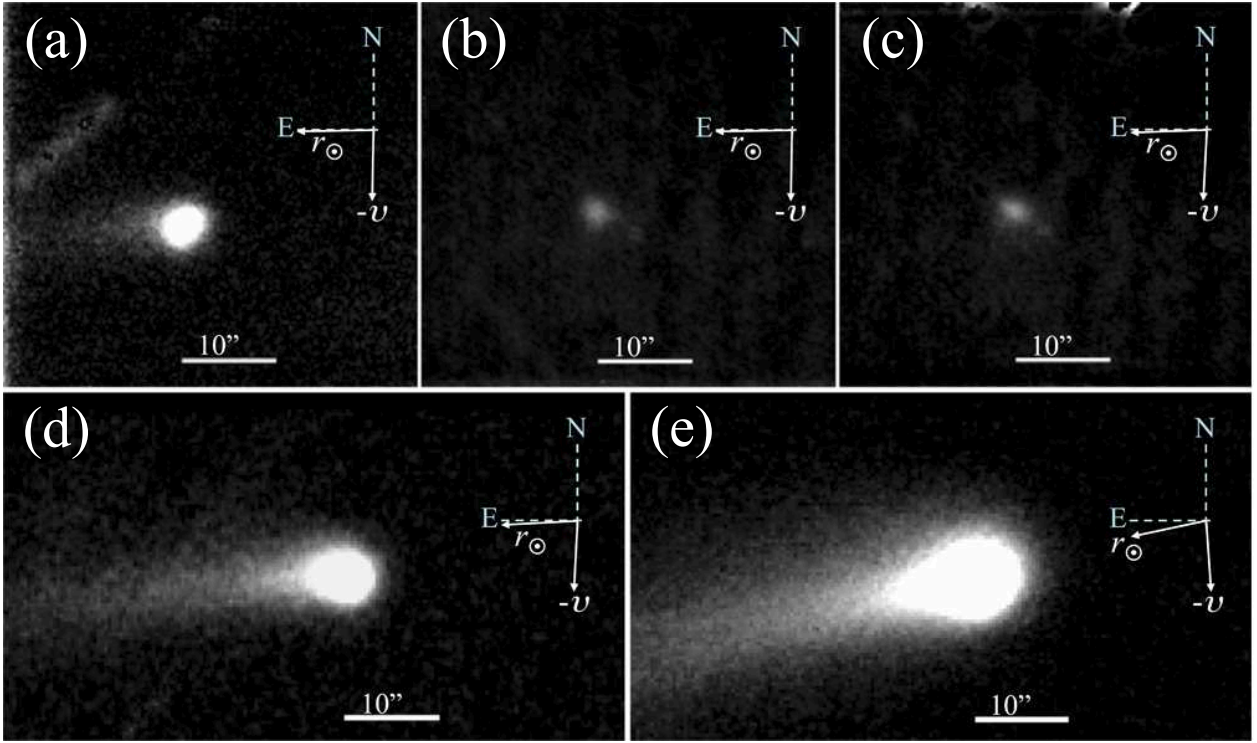}
  \caption{Optical and near-infrared ordinary images of 209P taken on
  (a) UT 2014 April 23 with HONIR in R$_\mathrm{C}$-band, 
  (b) UT 2014 April 23 with HONIR in J-band,   
  (c) UT 2014 May 1 with HONIR in J-band,
  (d) UT 2014 May 4 with MSI in R$_\mathrm{C}$-band, and
  (e) UT 2014 May 17 with MSI in R$_\mathrm{C}$-band.
  The comet nucleus is located at the center of each panel.
  The Celestial North is up and  Celestial East to the left. (a)--(c) have the field-of-view
  of 44\arcsec$\times$41\arcsec\ and (d)--(e) 67\arcsec$\times$38\arcsec.
  The Sun$\rightarrow$209P vector ($r_\odot$)
  and the negative heliocentric velocity vector ($-v$) are shown by arrows.}
 \label{fig:fig1}
\end{figure}

Observed raw data were processed in the standard manner for astronomical imaging data. All object frames were bias subtracted, flat fielded, and cosmic ray corrected using the IRAF reduction package. Figure \ref{fig:fig1} shows the processed images, which are produced  by summing up the reduced ordinary images. We found that the cometary tail was faint in the image on UT 2014 April 23 but became obvious in the R$_\mathrm{C}$-band image in May 2014.  The  tail extended between the Sun--comet vector and the negative heliocentric velocity vector, which is typical of cometary dust tails (rather than ion tails). On the contrary, the  tail was not detected in the J-band image due to the low S/N ratio. We noticed the comet was slightly stretched in the apparent direction of movement in the image taken with HONIR/Kanata because we could not employ a  comet-tracking mode with the Kanata telescope (theoretically, it stretched by 1.8\arcsec\ on May 23, 2.8\arcsec\ on May  1, and 3.7\arcsec\ on May 17, comparable to or slightly larger than the FWHM in HONIR images).

Imaging polarimetric data have often been analyzed to produce the two-dimensional polarization maps \citep{Furusho1999,Furusho2007,Jones2000,Hadamcik2013}. However, since the dust tail was too faint to permit us to make a map, we examined the spatial variation of the linear polarization degrees in the following manner: We first extracted individual source fluxes on the ordinary and extraordinary images by means of the IRAF {\it apphot} task. Each flux was numerically integrated over the desired circular aperture and the sky background was subtracted using the surrounding annulus  (i.e. aperture photometry). We set the aperture size ranging from one to five times the full width at half-maximum (FWHM, 2.6\arcsec--3.5\arcsec) of 209P. 
The extracted intensities are used to obtain the Stokes parameters normalized by the intensity, $Q/I$, and $U/I$, which are given by:

\begin{eqnarray}
\frac{Q}{I} = \left(1-\sqrt{\frac{I_{e,0} / I_{o,0}}{ I_{e,45} / I_{o,45}}}\right) / \left(1+\sqrt{\frac{ I_{e,0} / I_{o,0}}{I_{e,45} / I_{o,45}}}\right),
 \label{eq:Q/I}
\end{eqnarray}
\noindent
and

\begin{eqnarray}
\frac{U}{I} = \left(1-\sqrt{\frac{ I_{e,22.5} / I_{o,22.5}}{ I_{e,67.5} / I_{o,67.5} }}\right) / \left(1+\sqrt{\frac{ I_{e,22.5} / I_{o,22.5} }{ I_{e,67.5} / I_{o,67.5} }}\right),
 \label{eq:U/I}
\end{eqnarray}

\noindent
where $I_{o,\Psi}$ and $I_{e,\Psi}$ are the ordinary and extraordinary intensities at the half-wave plate angles $\Psi$ in degree \citep{Kawabata1999}. The degree of linear polarization ($P$) and the position angle of polarization ($\theta_P$) were calculated as 

\begin{eqnarray}
P = \sqrt{ \left(\frac{Q}{I}\right)^{2} + \left(\frac{U}{I}\right)^{2} },
 \label{eq:P}
\end{eqnarray}

\noindent
and

\begin{eqnarray}
\theta_P = \frac{1}{2} \tan^{-1} \left(\frac{U}{Q}\right),
 \label{eq:theta}
\end{eqnarray}

\noindent
 respectively \citep{Tinbergen1996}. For each night, we corrected for the instrumental polarization, the polarization bias, and the position angle zero-point using the results for the polarized and unpolarized standard stars. The polarization quantity ($P_r$)  and the position angle of the polarization plane ($\theta_r$) referring to the scattering plane \citep{Zellner1976} were expressed as the following:
 
 \begin{eqnarray}
P_r = P\cos(2\theta_r), 
 \label{eq:Pr}
\end{eqnarray}

\noindent
and

\begin{eqnarray}
\theta_r = \theta_P - (\phi \pm 90\arcdeg),
 \label{eq:theta}
\end{eqnarray}

\noindent
where $\phi$ is the position angle of the scattering plane (see Table \ref{tab:t1}), and  the sign inside the bracket is chosen to satisfy $0\arcdeg$ $\leq$ ($\phi$ $\pm$ $90\arcdeg$) $\leq$ $180\arcdeg$ \citep{Chernova1993}. In Table \ref{tab:t2}, we summarize the results of the polarization degrees ($P$ and $P_r$) and the position angles ($\theta_P$ and $\theta_r$), which are derived from the multiple data sets at each night.
 %\textcolor{red}{We used data of XXX as the strongly polarized standard star and XXX as the unpolarized standard star with HONIR, whereas XXX as the strongly polarized standard star and XXX as the unpolarized standard star with MSI, respectively.}

\subsection{Nuclear Polarization}
\begin{figure}
 \epsscale{1.0}
   \plotone{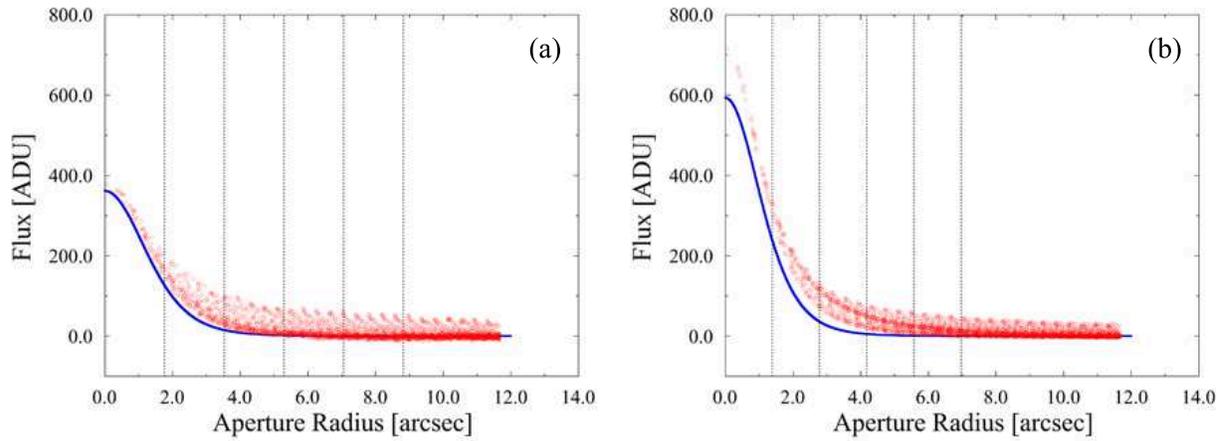}
  \caption{Radial profiles in the ordinary images from observations made on (a) UT 2014 May 4 and (b) UT 2014 May 19 using MSI in R$_\mathrm{C}$-band.  
  The open circles show the extracted data points as the radial profile of 209P.
  The solid line shows the point-spread-function of the star scaled to the nuclear flux.
  The vertical dotted lines correspond to 1--5$\times$FWHM of the comet. 
  From the comparison, it is obvious the nuclear flux is dominated at smaller distance (within $\sim$2$\times$FWHM) while coma flux dominates at a greater distance (beyond $\sim$2$\times$FWHM)
  }\label{fig:f2} 
\end{figure}

From this point on, we tried to distinguish the nuclear flux from the coma flux. In general, the optical depth of most cometary comae is small relative to the nucleus. It is unclear for the flux contrast between the nucleus and the coma \citep{Lamy2004}. We chose the data from two nights (UT 2014 May 4 and 19) obtained with Pirka to allow the use of the precise non-sidereal tracking. The radial profiles of 209P were obtained from the co-added frames, which used only ordinary intensities at the four half-wave plate angles. Stellar radial profiles were defined from the standard star images in the same night. These data were converted into the one-dimensional azimuthally averaged surface brightness by the {\it pradprof} task in IRAF. The profile parameters were determined by the best fit model with the Moffat PSF fitting function \citep{Moffat1969}. The reduced magnitudes of the nucleus on these two nights, $R$=19.92$\pm$0.26 mag and $R$=20.02$\pm$0.26 mag, were calculated using an empirical formula, $R$=16.24+0.04$\alpha$ \citep{Ishiguro2015}. Figure \ref{fig:f2} shows the radial profiles 209P (open circles) together with nucleus model (solid line) on these two nights. It is clear that the observed fluxes are dominated by the nucleus at the radial distance within $\sim$2$\times$FWHM, while by the coma beyond $\sim$2$\times$FWHM.

\begin{figure}
 \epsscale{1.0}
   \plotone{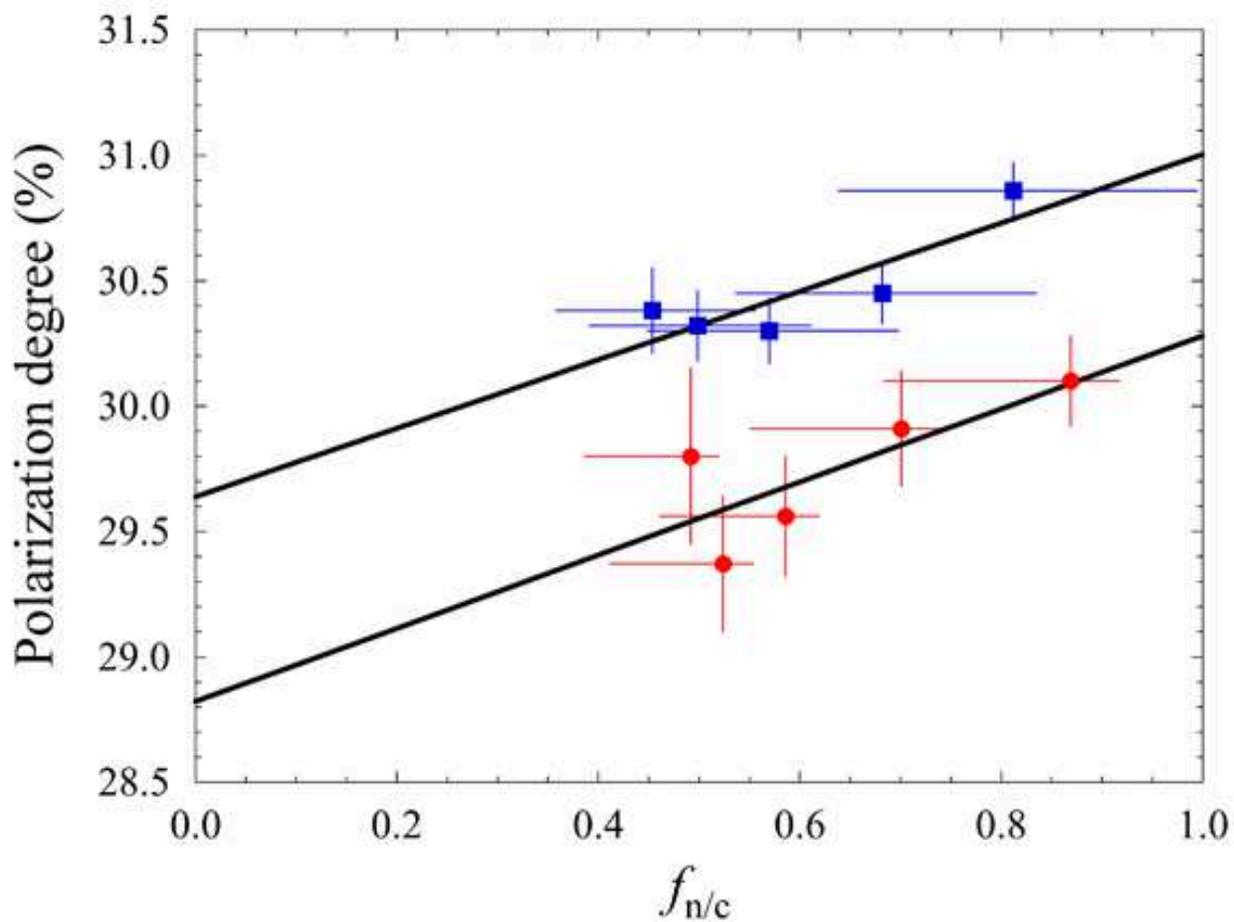}
  \caption{Polarization degree ($P$) with respect to ratio of the nuclear flux to the observed flux $f_{n/c}$ taken on UT 2014 May 4 (filled circles) and 19 (filled squares). 
  The solid lines are fitted ones for $P$--$f_{n/c}$ plots on each night.  
  The intercept along the vertical axis (i.e. $f_{n/c}=0$) asymptotically approaches the coma polarization degree whereas the intersection with $f_{n/c}$=1 approximately equals the polarization degree of the nucleus.
  }\label{fig:f3} 
\end{figure}

To distinguish the nuclear polarization degrees from the coma polarization degree, we made a plot of the polarization degrees taken with different aperture sizes with respect to the ratio of the modelled nuclear flux to the observed flux, $f_{n/c}$ (Figure \ref{fig:f3}). We considered the error associated not only with the signal-to-noise ratio and the uncertainty in the sky background determination, but also with the uncertainty of the nuclear magnitude model \citep[i.e. 0.26 mag,][]{Ishiguro2015}. In Figure \ref{fig:f3}, the data with smaller apertures are plotted in the upper right, while the data with larger apertures are plotted in the lower left. This trend implies that the cometary nucleus is more polarized than the coma. In principal, the $P$-intercept (i.e. $f_{n/c}$=0) is supposed to approach the coma polarization degree while the intersection with $f_{n/c}$=1 approaches the nuclear polarization degree. We thus assumed that the data points in Figure \ref{fig:f3} can be fitted with a simple linear function, $P$ = ($P_c-P_n$) $\times$ $f_{n/c}$ + $P_n$, where $P_n$ and $P_c$ are polarization degrees of the nucleus and coma, respectively, and derived these values by the method of least squares.

%%%%%%%%%%%%%%%%%
%     SECTION 3: RESULTS      %
%%%%%%%%%%%%%%%%%

\section{Results}
As the result of fitting, we obtained the polarization degree $P_n$=30.3$^{+1.4}_{-0.9}$\% at $\alpha$=92.2\arcdeg\
and $P_n$=31.0$^{+1.0}_{-0.8}$\% at $\alpha$=99.5\arcdeg\ for the nucleus, and $P_c$=28.8$^{+0.4}_{-0.4}$\%
at $\alpha$=92.2\arcdeg\ and$P_c$=29.6$^{+0.3}_{-0.3}$\% at $\alpha$=99.5\arcdeg\ for the coma.
A somewhat surprising result may be derived as the polarization degree for the nucleus.

% Maximum polarization 
\subsection{Positive branch of polarization}
In Table \ref{tab:t2}, the position angle of the strongest electric vectors, $\theta_P$, is in good agreement with those of normal vector to the scattering plane projected on the celestial plane, $\theta_\perp$ (see Table \ref{tab:t1}) to an accuracy of a few degrees. For a spherical body, $\theta_P$ is supposed to align to the normal vector of the scattering plane, and therefore, $\theta_P$=$\theta_\perp$.  
Although the radar image of 209P showed an irregular shape \citep{Howell2014}, such a regional irregularity may be smoothed out by integrating the flux through the disk. 
These high polarizations for nucleus indicate a light scattering feature on an object a few kilometers in size \citep[2.5 km$\times$3.2 km,][]{Ishiguro2015}.
Since the polarization degree for an airless body is primarily controlled by the surface albedos,  the low albedo of the nucleus \citep[the geometric albedo $P_v\sim$0.05,][]{Ishiguro2015}would result in the high polarization degree. However, we are unable to confirm this because of a lack of precedent for such dark airless bodies (typically $P_v<$0.1). 

The following studies were shown potential for becoming the high polarization degree ($\sim$30\%) of the dark bodies around $\alpha$=95\arcdeg.
There are studies of near-nuclear polarization for 2P/Encke \citep[$P_v$=0.05$\pm$0.02,][]{Fernandez2000}.  
High and low polarizations have been reported to $P$=34.9\%$\pm$1.7\% at $\alpha$=94.6\arcdeg\ ($P$=39.9\%$\pm$2.9\% when corrected for the NH$_\mathrm{2}$ contamination) in the red domain (662/5.9nm) \citep{Jorkers2005} and $P$=11\%$\pm$3\% at $\alpha$=92.9\arcdeg\ in the green domain (526/5.6nm) \citep{Jewitt2004} respectively. 
Considering the contamination by the gaseous species,  both studies were measured through the narrow band filter for the same appearance. 
Note the difference that only the spectral gradient is incompletely explained as pointed out by \citet{Hadamcik2009}.
Polarization of dark airless bodies at the high phase angle were obtained $P$=24.5\%$\pm$4\% at $\alpha$=81\arcdeg\ for the Martian satellites Phobos \citep[$P_v$=0.07,][]{Zellner1974a} and $P$=22\%$\pm$4\% at $\alpha$=74\arcdeg\ for Deimos\citep[$P_v$=0.07,][]{Thomas1996} in the orange domain (570nm),  while these measurements by Mariner-9 may contain significant systematic errors \citep{Noland1973}. 
Thus, more observations of dark asteroids are needed to investigate the positive polarization at the higher phase angle.

\begin{figure}
 \epsscale{1.0}
   \plotone{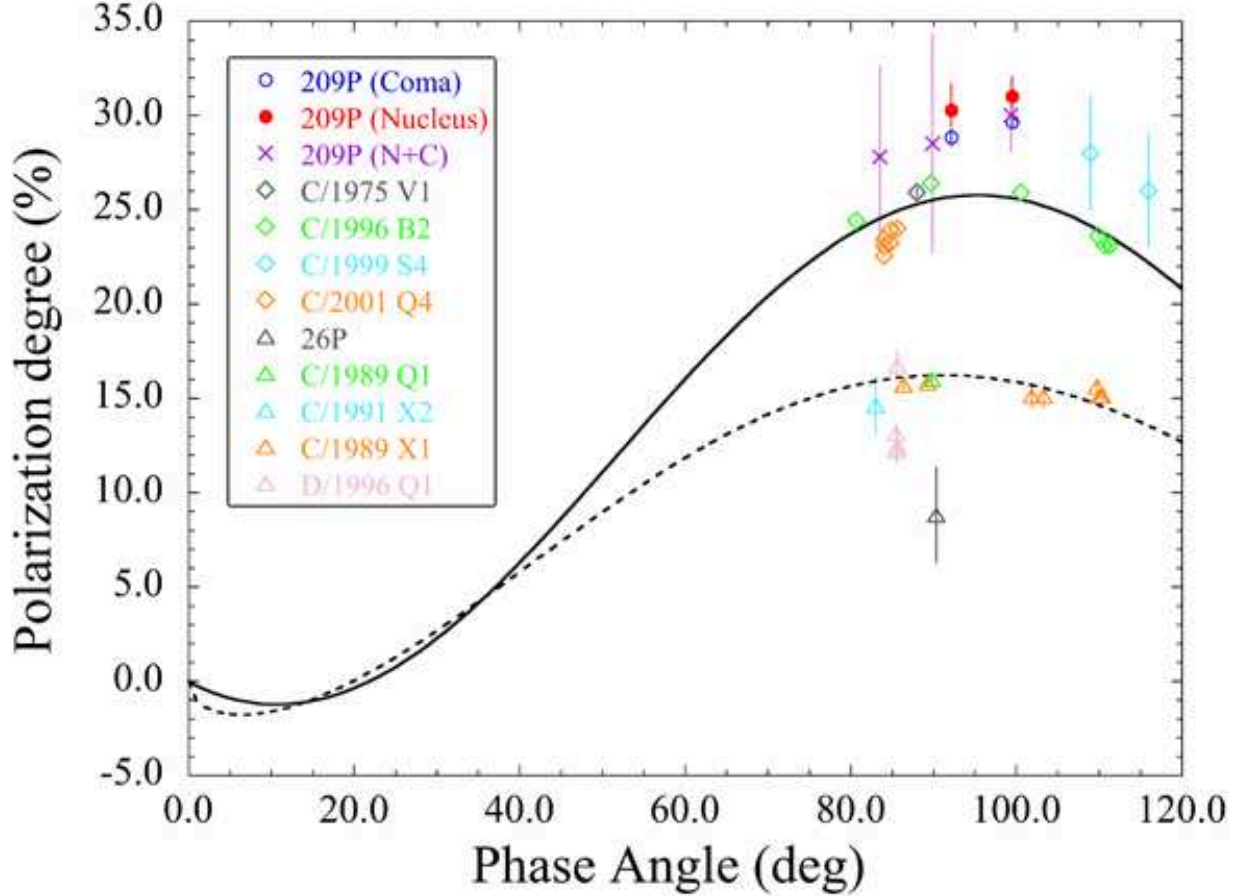}
  \caption{Phase angle dependence of polarization for 209P nucleus (filled circles) and coma (open circles) taken with MSI. 
Regarding data taken with HONIR (indicated by crosses), we adopted blended signals of nucleus and coma with relatively large aperture (3$\times$FWHM) due to inadequate tracking of the telescope. 
For comparison, we show polarization degrees of other comets in R-band region (about 650$\pm$50 nm), which are mostly attributed to light scattered by dust particles in comae.  
Two fit lines with the Lumme and Muinonen function, high $P_\mathrm{max}$(solid line and open diamonds) and low $P_\mathrm{max}$(dashed line and open triangles) comprise the PDS archive\citep{Kiselev1978,Michalsky1981,Myers1985,Kikuchi1987,Kikuchi1989,Eaton1988,Eaton1992,Rosenbush1994,Levasseur1996,Hadamcik2003a,Hadamcik2003b,Joshi2003,Kikuchi2006,Ganesh2009}}. 
 \label{fig:f4} 
\end{figure}

For the coma, such positive polarization with a maximum of $P_\mathrm{max}$ = 25--30$\%$ around $\alpha$=95\arcdeg\ in the red domain was reported in previous studies of the high $P_\mathrm{max}$ comets (see Figure \ref{fig:f4}). Our measurements for the coma are consistent with them. We found from HONIR measurements that the wavelength dependence of the polarization for the mixture of the nucleus  and coma was negligibly small, that is,  $P$=30\%$\pm$3\% in R$_\mathrm{C}$-band and $P$=27\%$\pm$5\% in J-band at $\alpha$=89.9\arcdeg\ with the aperture size of 1--3$\times$FWHM, although the error of the measurement was not good enough to compare with previous studies.
%
%These data suggest that  the polarization degree at more than $\alpha$=90\arcdeg may stand at nearby $\sim$ 30\% in the red domain. 
%Such positive polarization with a maximum of $P_\mathrm{max}$ = 30$\%$ around $\alpha$=95\arcdeg in the red domain were reported in previous studies of the high $P_\mathrm{max}$ comets (see Figure \ref{fig:f4} ). Our measurements for the coma are consistent with them.
%As we mentioned above, the polarization degree of the nucleus is slightly higher than that of the coma. 
%We found that the polarization degree of the nucleus is significantly higher than that of a previously studied comet, 2P/Encke \citep[$P$=11\%$\pm$3\% at $\alpha$=92.9\arcdeg][]{Jewitt2004}.
%Since the polarization degree is primarily controlled by the surface albedos but the albedo of 2P/Encke \citep[0.05$\pm$0.02,][]{Fernandez2000} is almost equivalent to that of 209P \citep[$\sim$0.05,][]{Ishiguro2015}, the other physical conditions of these bodies such as the regolith sizes may differ,  although we are unable to discuss these differences further because physical information of these two short-periodic comets is limited.

% Phase angle dependence of P
\subsection{Phase angle dependence of $P$}
Figure \ref{fig:f4} shows the phase angle dependence of the polarization degrees of the nucleus and coma. 
For comparison, we show the previous studies for cometary comae at phase angles larger than about 80\arcdeg\ and two red continuum lines through the use of the database of comet polarimetry \citep{Kiselev2005}. Our data shows a moderate increase from 83.5\arcdeg\ to 99.5\arcdeg\ and no clear peak. 
There are known to be two classes of comets in terms of the maximum polarization $P_\mathrm{max}$, so-called 'high-polarization comets' and 'low-polarization comets'. 
\citet{Levasseur1996} reported that these two classes of comets have similar polarization properties at $\alpha<$40\arcdeg\ but are bifurcated at larger phase angles. 
The high-polarization comets have the peak polarization $P_\mathrm{max}\sim$25\%--28\% around $\alpha$=90\arcdeg--100\arcdeg, while the low-polarization comets have $P_\mathrm{max}\sim$15\% around the same phase angle. 
In addition, \citet{Levasseur1996} described the high-polarization comets as tending to be classified as dusty comets, while the low-polarization comets tend to be classified as gassy comets based on optical spectra.
However, the latter case was evaluated as a misleading definition due to the gas contaminations through the broadband filter \citep{Kiselev2001,Kiselev2004, Jewitt2004}. 
In the near-nucleus region, the polarizations of some low-polarization comets reach to those of high-polarization comets \citep{Kolokolova2007}.

\section{Discussion}
\label{sec:discussion}
% contamination of gas
\subsection{Contamination of gas}
\begin{figure}
 \epsscale{1.0}
   \plotone{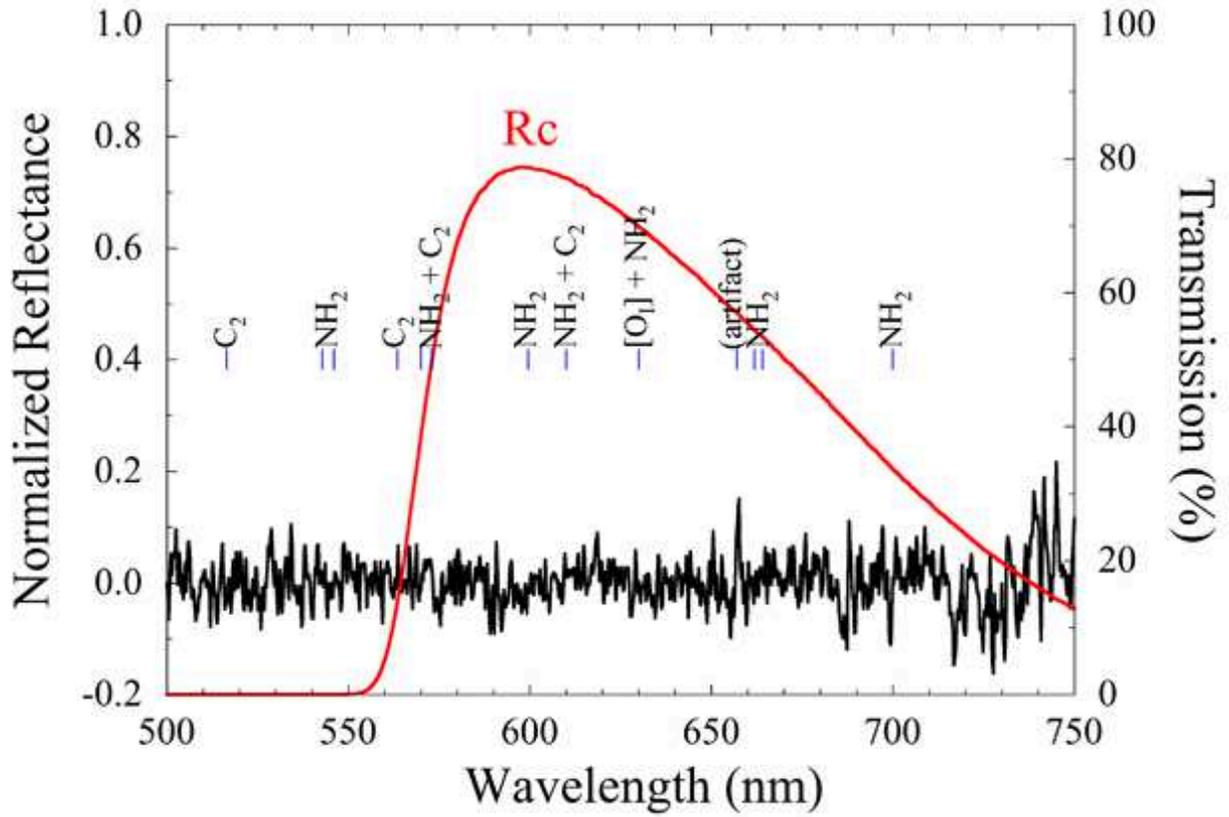}
  \caption{Continuum-subtracted spectrum of 209P taken at the Nishi-Harima Astronomical Observatory with MALLS. Major emission lines are labelled and the peak at 658 nm is evaluated as a calibration artifact. The transmission curve of our R$_\mathrm{C}$ filter is indicated too. }
\label{fig:f5} 
\end{figure}

First of all, we would like to discuss the contamination of polarization degree by gas emissions. 
The observed polarization degrees of comets are known to be occasionally contaminated by gas emissions that reduce the observed polarization degree. 
In fact, gas-rich comets tend to display low polarization degrees probably because of the depolarization of the gas emission \citep{Kiselev2001}. 
However, we found that the observed spectra of the 209P coma showed no detectable gas emission in the wavelength range in which we deduced the polarization degree. 
Figure \ref{fig:f5} shows the optical spectra on UT 2014 May 27, only about a week after our polarimetric run, taken at the Nishi-Harima Astronomical Observatory with the Medium and Low--dispersion Long--slit Spectrograph (MALLS) attached to the Nayuta 2.0--m telescope.  
We employed the lowest-resolution grating (150 line mm$^{-1}$), 8\arcsec--slit, and the order cut filter of WG320, ensuring the available spectrum at the wavelength 500--730 nm with the spectral resolution $R\sim$90. 
Note that the continuum slope might be affected by insufficient calibration of the flux with the standard star taken at slightly different airmass.
We normalized the obtained reflectance spectrum at 550 nm and subtracted the continuum by the spline interpolation.
Figure \ref{fig:f5} shows the continuum-subtracted spectrum with the typical emission lines labelled \citep{Brown1996}. 

There are clearly no detectable emission lines in the coverage of our R$_\mathrm{C}$ filter, dispelling the gas contamination. 
Similar results are reported through spectroscopic observation on UT 2014 April 9 that a spectrum in the 350--600 nm range reveals no CN, C$_2$, or C$_3$ emission features \citep{Schleicher2014}. 
Although there might be time variation in the intensity of gas emission, we conjecture that the variation would be small in our polarimetric data because the R$_\mathrm{C}$-band monitoring magnitudes did not change over this period \citep{Ishiguro2015}.

We distinguished the polarization of the nucleus $P_n$=30--31\% from that of the coma $P_c$=29--30\% taking account of the signal with different apertures (see Section 2.3). 
These two values are not necessarily the same, because we observe the scattered light from the coma dust grains in the former, while we observe the reflected light from a kilometer-sized body in the latter. 
As shown in Figure \ref{fig:f4}, a comparison between observations performed in other comets indicates a lower trend for the coma polarizations of 209P.
We may  nevertheless underestimate the polarization degrees for the coma due to a very slight gas contamination. 

\subsection{Estimation of maximum polarization $P_\mathrm{max}$}
The polarization of the nucleus ($P_n$) seems to show higher than the coma ($P_c$) but the differences are insignificant to the accuracy of our measurement. 
Figure \ref{fig:f6} compares the polarization degrees of the nucleus with those of asteroids at phase angles larger than about 60\arcdeg\ from the asteroid polarimetric database \citep{Lupishko2014}.  All asteroids in the database show polarization degrees remarkably lower than 209P. 
The phase-polarization curve of (4179) Tautatis is well studied among these asteroids.
\citet{Ishiguro1997} determined the phase angle of the maximum polarization at $\alpha_\mathrm{max}$=107\arcdeg$\pm$10\arcdeg.
Although we cannot determine $\alpha_\mathrm{max}$ of 209P with our measurement, we conjecture that our polarimetric measurement was performed around the maximum polarization degree because both the S-type asteroid and cometary coma show the maximum values around the phase angle we measured.
For a better understanding of the polarization properties of the nucleus and coma dust at large phase angles, we derived the $P_\mathrm{max}$ using the Lumme and Muinonen function \citep{Goidet-Devel1995,Penttila2005}:

\begin{eqnarray}
P_n=b\left(\sin \alpha\right)^{c_1} \left( \cos\left(0.5\alpha\right)\right)^{c_2} \sin\left(\alpha-\alpha_0\right),
 \label{eq:Pmax}
\end{eqnarray}

\noindent
where $b$, $c_{1}$, $c_{2}$ and $\alpha_0$ are constant parameters. 
According to \citet{Penttila2005}, parameters $b$ and $\alpha_0$ affect the amplitude and slope of polarization and the inversion angle, and the other two parameters, the powers  $c_{1}$ and  $c_{2}$, control shapes the phase curve. 
Because the phase angle coverage of our data is limited, we first fit the phase-polarization relations for comets or asteroids, and then derived $P_\mathrm{max}$ of 209P.  
Therefore, assuming that the nucleus was essentially identical to low-albedo asteroids($P_v<$0.1), we deduced $P_\mathrm{max}$ of the nucleus is 30.8\% at $\alpha_\mathrm{max}$ of 96\arcdeg\ with the fitted parameters $b$=34.81\%, $c_{1}$=0.7838, $c_{2}$=0.2400 and $\alpha_0$=17.88\arcdeg. For the coma, assuming that the fitted parameters $c_{1}$=0.7904, $c_{2}$=0.2418 and  $\alpha_0$=21.59\arcdeg\ for high-polarization comets and keeping $b$ as a free parameter, we deduced $P_\mathrm{max}$ of the coma is 29.6\% with $\alpha_\mathrm{max}$=98\arcdeg.

Our $P_\mathrm{max}$ values are higher polarization than other high-polarization comets and known airless bodies.  
It was suggetsed that the presence of jets and arcs in the ambient coma correlated with regions of higher polarization \citep{Tozzi1997,Furusho1999,Hadamcik1997}. 
However, according to \citet{Ishiguro2015}, such structures were not identified for 209P.
Trends are shown as that are similar to the case with cometary nuclei split into fragments (e.g. C/1999 S4 \citep{Hadamcik2003c} and 73P \citep{Bonev2008}),  whereas 209P is obvious without such an outburst. Therefore the correlation between cometary activity and polarization excess are considered to be unsuitable for 209P.
Some physical properties such as grain size distribution and porosity \citep{Hanner2003,Hadamcik2002,Hadamcik2006} may enhance the polarization degree of 209P, although we have no observational evidence to study these properties.

% comparison with asteroids
\subsection{Comparison with asteroids and slope-albedo relationship}
\begin{figure}
 \epsscale{1.0}
   \plotone{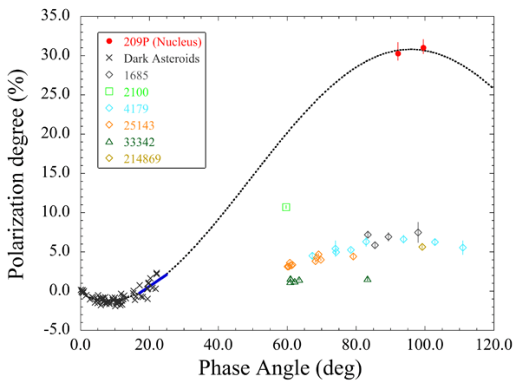}
  \caption{Comparison of phase angle dependence of polarization of 209P at R$_\mathrm{C}$--band, reported in this paper, with those of near--Earth asteroids, 1685 Toro \citep{Kiselev1990}, 4179 Toutatis \citep{Ishiguro1997,Mukai1997}, 2100 Ra-Shalon \citep{Kiselev1999},  25143 Itokawa \citep{Cellino2005}, 33342 1998 WT$_\mathrm{24}$\citep{Kiselev2002}, and 214869 2007 PA$_\mathrm{8}$\citep{Fornasier2015}.  The dotted line is produced  by means of the Lumme and Muinonen function between 209P nucleus and the dark asteroids from the PDS archive \citep{Zellner1974b,Belskaya1987,Nakayama2000,Belskaya2005,Fornasier2006,Belskaya2009,Bagnulo2010,Canada-Assandri2012,Gil-Hutton2012,Lupishko2014}.
  The solid line shows a slope of the linear region of the phase-polarization curve.}
  %Solid and dashed lines are shown the Lumme and Muinonen function for 209P nucleus and coma respectively.
\label{fig:f6} 
\end{figure}

Asteroids show a similar trend of the phase-polarization relation having the maximum polarization at $\alpha$=90\arcdeg--120\arcdeg\ \citep{Ishiguro1997,Kiselev1999} but the polarization degrees of these asteroids are significantly lower than that of the 209P nucleus (see Figure \ref{fig:f6}).
Note that only a few asteroids including (1685) Toro, (4179) Toutatis (214869) 2007 PA$_\mathrm{8}$ were observed at large phase angles and they are classified as S-type asteroids.  
The S-type asteroids show very similar phase-polarization relations. 
Unique data concerning E-type asteroids, which was obtained from (33342) 1998 WT$_\mathrm{24}$, indicated $P_\mathrm{max}$ at $\alpha$=72\arcdeg\ \citep{Kiselev2002}.
%There S-type asteroids show very similar phase --polarization relation, displaying $P_\mathrm{max}$=7.0$\pm$0.2\% \citep{Ishiguro1997}. 
The linear polarization of (2100) Ra-Shalom (C-type) was measured up to a moderate phase angle ($\alpha$=50\arcdeg). 
The C-type asteroid shows large $P$ at $\alpha\gtrsim$20\arcdeg, although $P_\mathrm{max}$ for C-type asteroids has not been determined yet.  
Since S-type asteroids generally have albedos larger than those of C-type asteroids and E-type asteroids have albedos larger than others, 
it seems that $P_\mathrm{max}$ for asteroids would have a correlation between albedo and $P_\mathrm{max}$.

A slope of the linear region at the inversion angle $h$ (as the solid line in Figure \ref{fig:f6} shows) was derived as 0.30\%/\arcdeg$\pm$0.01\%/\arcdeg\ from the phase-polarization curve. This polarimetric slope $h$ was slightly steeper than the mean value of C-type objects \citep[0.282\%/\arcdeg$\pm$0.025\%/\arcdeg,][]{Gil-Hutton2012}.
The relation between the geometric albedo $P_v$ and the slope $h$ has been found from the studies of the scattering properties \citep{Zellner1974b,Dollfus1989,Cellino1999}, and is the following:
\begin{equation}
	 log\ P_v=C_1\ log\ h + C_2~~,
 \label{eq:Pv-h}
\end{equation}
\noindent
where $C_1$ and $C_2$ are the constants.
These recent values were presented as $C_1$=--1.207$\pm$0.067 and $C_2$=--1.892$\pm$0.141 by \citet{Masiero2012}.
The geometric albedo for the nucleus of 209P is calculated 0.05$^{+0.03}_{-0.02}$ on these constraints, which are consistent findings for \citet{Ishiguro2015} and 0.049$\pm$0.020 as the average albedo of the potential dormant comets \citep{Kim2014}.
%
%Therefore, assuming that the fitted parameters $c_{1}$=0.7904, $c_{2}$ = 0.2418 and  $\alpha_0$=21.59\arcdeg for high-polarization comets and keeping $b$ as a free parameter, we deduced $P_\mathrm{max}$ of the nucleus and the coma are 30.8\% and 29.6\%, respectively with the common $\alpha_\mathrm{max}$ value of 98\arcdeg.
%
%There are known to be two classes of comets in terms of the maximum polarization $P_\mathrm{max}$, so called 'high-polarization comets' and 'low-polarization comets' (see Figure \ref{fig:f4}). 
%\citet{Levasseur1996} reported that these two classes of comets have similar polarization properties at $\alpha<$40\arcdeg\ but bifurcated at larger phase angles. 
%The high-polarization comets have the peak polarization $P_\mathrm{max}\sim$25\%--28\% around $\alpha$=90\arcdeg--100\arcdeg, 
%while the low-polarization comets have $P_\mathrm{max}\sim$15\% around the same phase angle. 
%In addition, \citet{Levasseur1996} described the high-polarization comets as tending to be classified as dusty comets, while the low-polarization comets tend to be classified as gassy comets based on optical spectra. 
%The comet 209P, which is most likely dust rich comet on the ground that it did not display the optical gas emission in the spectrum, has $P_\mathrm{max}$ of the coma comparable to those of high-polarization comets. The trend is in accordance with the classification in \citet{Levasseur1996} classification. 

\subsection{$P_\mathrm{max}$--albedo relationship}
% Umov law
This $P_\mathrm{max}$--albedo relationship has long been known as the Umov law. Since the effect is observed on the Moon, Mercury and Mars, the Umov law could essentially be applicable to asteroids and comets. 
The law is described as \citep{Dollfus1998}:
\begin{equation}
P_\mathrm{max}= C_3 \left(d\right) P_v^{-C_4}~~,
 \label{eq:Umov}
\end{equation}
\noindent
where $P_v$ is the geometric albedo. $C_3$ is a constant, originally proposed $C_4$=1 by Umov.
Also, $C_3\left(d\right)$ depends a grain size $d$ of the regolith.

\begin{figure}
 \epsscale{1.0}
   \plotone{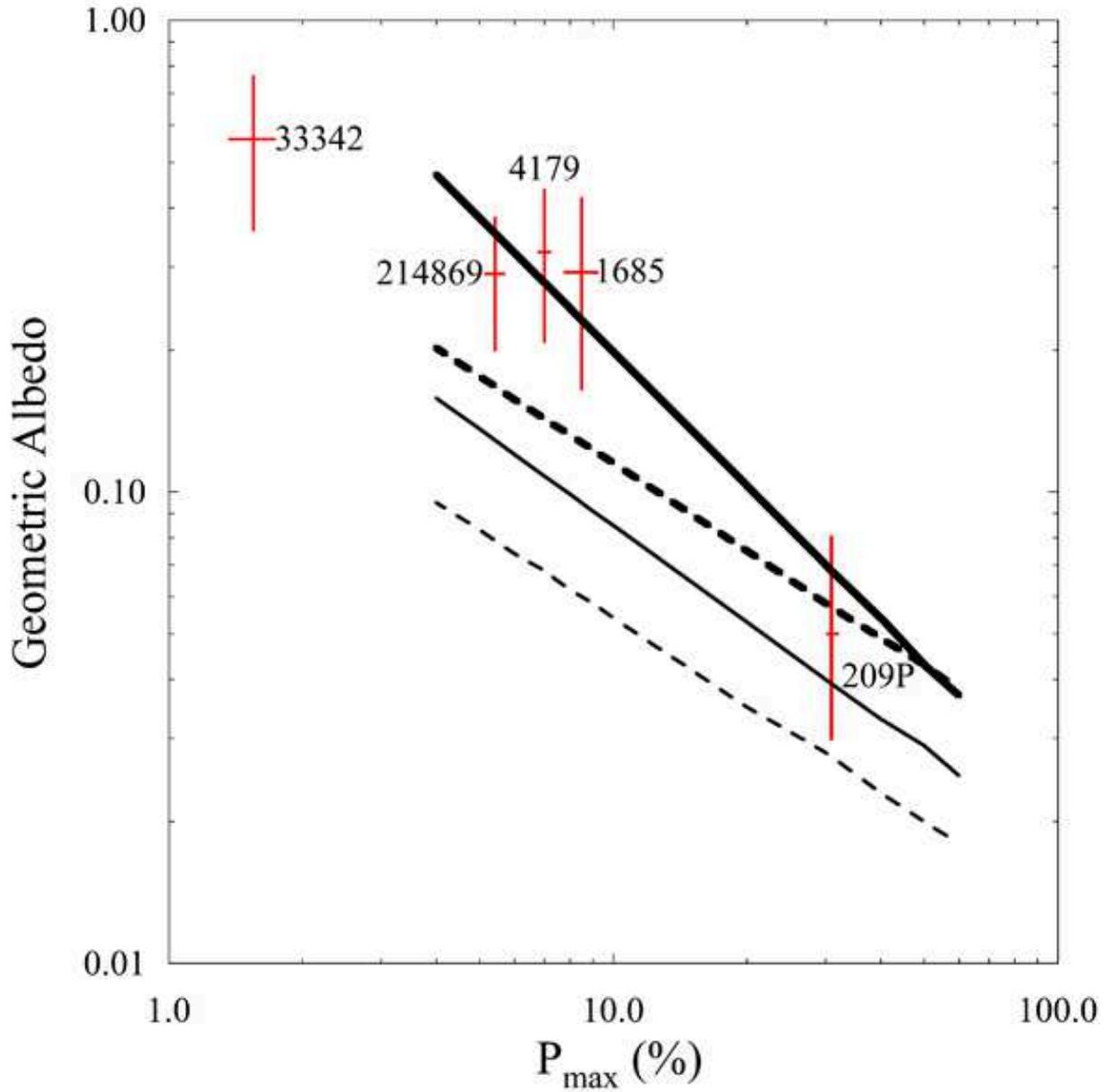}
  \caption{Geometric albedo with respect to $P_\mathrm{max}$ for (1685) Toro, (4179) Toutatis (214869) 2007 PA$_\mathrm{8}$ (S-type), (33342) 1998 WT$_\mathrm{24}$ (E-type), and 209P. Albedo -- $P_\mathrm{max}$ relations based on two models shown by solid \citep{Dollfus1998} and dashed lines \citep{Shevchenko1995} are plotted as typical surface grain sizes: coarse grain(160 $\micron$) and fine grain(1-4 $\micron$). The bold lines are also indicators for the coarse grain.
    }\label{fig:f7} 
\end{figure}

\citet{Dollfus1998} investigated how $C_3$ and $C_4$ change according to the grain size. 
Figure \ref{fig:f7} shows the albedo v.s. $P_\mathrm{max}$ plot of 209P together with the Umov function modified by \citet{Dollfus1998} and \citet{Shevchenko1995}. 
%The geometric albedo of 209P was unknown, so we adopted 0.049$\pm$0.020 as the average albedo of the potential dormant comets \citep{Kim2014}.
The geometric albedo of 209P was unknown,  so we adopted 0.05$^{+0.03}_{-0.02}$ as the likely estimate from Equation (\ref{eq:Pv-h}).
We also plotted the data of (1685) Toro, (4179) Tautatis,  (33342) 1998 WT$_\mathrm{24}$, (214869) 2007 PA$_\mathrm{8}$ referring to the geometric albedo of 0.29$\pm$0.13 \citep{Masiero2012}, 0.32$\pm$0.11 \citep{Kraemer2005}, 0.56$\pm$0.20 \citep{Harris2007}, and 0.29$\pm$0.09 respectively.
It is likely the albedo--$P_\mathrm{max}$ relation can be explained by the Umov law. 
Although the errors of albedos are too large to enable the regolith sizes on these bodies to be determined, we can safely say that our 209P data favours large ($\gg$1 \micron) particle sizes. 
Submicron-sized particles were probably well accommodated by gas outflow and lost over time by repeated comet activities.

\section{Summary}
\label{sec:summary}
We made polarimetric observations of the highly dormant Jupiter-Family Comet, 209P, at R$_\mathrm{C}$ and J--band during the perihelion passage in 2014 and found the following:
\begin{enumerate}
\item{No significant difference was found in R$_\mathrm{C}$- and J-band.}
\item{Both nucleus and dust coma show high linear polarization degree, that is, $P_n$=30.3$^{+1.3}_{-0.9}$\% at $\alpha$=92.2\arcdeg\ and $P_n$=31.0$^{+1.0}_{-0.7}$\% at $\alpha$=99.5\arcdeg\ for the nucleus, and $P_c$=28.8$^{+0.4}_{-0.4}$\% at $\alpha$=92.2\arcdeg\ and 29.6$^{+0.3}_{-0.3}$\% at $\alpha$=99.5\arcdeg\ for the coma.}
%\item{Both nucleus and dust coma show high linear polarization degree, that is, $P_n$=30.3$^{+1.3}_{-0.9}$\% at $\alpha$=92.2\arcdeg\ and $P_n$=31.0$^{+1.0}_{-0.7}$\% at $\alpha$=99.5\arcdeg\ for the nucleus, and $P_c$=28.8$^{+0.4}_{-0.4}$\% at $\alpha$=92.2\arcdeg\ and 29.6$^{+0.3}_{-0.3}$\% at $\alpha$=99.5\arcdeg\ for the coma.}
\item{We detected no significant variation in $P$ at the phase angle coverage, suggesting that the obtained $P$ is nearly at maximum in the phase-polarization curves. We deduced $P_\mathrm{max}$=30.8\% (nucleus) and 29.6\% (coma), respectively.}
\item{High $P_\mathrm{max}$ value of the dust coma is consistent with a polarization classification scheme described by \citet{Levasseur1996}.}
\item{We employed an empirical function for relating $P_\mathrm{max}$ of the nucleus to the albedo, 
and found that we obtained a good estimate of the albedo when we assumed the effective size of the regolith particles of $\approx$1--100\micron.} 
The Umov law seems to work on the cometary nuclei as well as the well-studied lunar surface.

\end{enumerate}

\vspace{1cm}
{\bf Acknowledgments}\\
We thank an anonymous referee for helpful comments.
This work was supported by the Optical and Near-infrared Astronomy Inter-University Cooperation Program 
and Grants-in-Aid for Scientific Research (23340048, 24000004, 24244014, and 24840031)
from the Ministry of Education, Culture, Sports, Science and Technology of Japan.
MI was supported by a National Research Foundation of Korea (NRF) grant funded by the Korean government (MEST) (No. 2012R1A4A1028713). 
SH was supported by the Space Plasma Laboratory, ISAS, JAXA (the Hypervelocity Impact Facility).

%%
%% Table 1
\begin{deluxetable}{ccrrrrrrrrrl} 
\tablecolumns{9} 
\tablewidth{0pc} 
\tablecaption{Observational circumstances.} 
\tablehead{ 
%\colhead{}    &  \multicolumn{3}{c}{Non-shell Stars} &   \colhead{}   & 
%\multicolumn{2}{c}
%\cline{2-4} \cline{6-8} \\ 
\colhead{Date} & \colhead{UT} & \colhead{Filter}  & \colhead{$Exp^{(a)}$} & \colhead{$N^{(b)}$} & \colhead{r$^{(c)}$} & \colhead{$\Delta^{(d)}$} & \colhead{$\alpha^{(e)}$} & \colhead{$\theta_\perp^{(f)}$} &  \colhead{$\phi^{(g)}$} & \colhead{Inst.$^{(h)}$}
%\\ \colhead{} & \colhead{}   & \colhead{}        & \colhead{[sec]}& \colhead{[AU]} & \colhead{[AU]} & \colhead{[$\arcdeg$]} & \colhead{[$\arcdeg$]} & \colhead{} 
}
\startdata
2014/04/23& 12:21-13:43 &R$_\mathrm{C}$& 132 &12 & 0.987 & 0.335 & 83.5 & 1.3 & 91.3 &HONIR\\
2014/04/23& 12:21-13:43 &J& 120 &12 & 0.987 & 0.335 & 83.5 & 1.3  & 91.3 &HONIR\\
2014/05/01& 13:01-13:48 &R$_\mathrm{C}$& 132 &12 & 0.972 & 0.268 & 89.9 & 2.4  & 92.4 &HONIR\\
2014/05/01& 13:01-13:48 &J& 120 &12 & 0.972 & 0.268 & 89.9 & 2.4  & 92.4 &HONIR\\
2014/05/04& 13:04-14:31 &R$_\mathrm{C}$& 180 & 24 & 0.970 & 0.242 & 92.2 & 3.2 & 93.2 &MSI\\
2014/05/17& 12:23-14:29 &R$_\mathrm{C}$& 45 & 54 & 0.983 & 0.127 & 99.4 & 10.2& 100.2 &HONIR\\
2014/05/19& 11:23-12:09 &R$_\mathrm{C}$& 60 & 36 & 0.988 & 0.110 & 99.5 & 12.2 & 102.2 &MSI\\
\enddata 
   \tablenotetext{(a)}{Individual effective exposure time in seconds.}
   \tablenotetext{(b)}{Number of exposures.}
   \tablenotetext{(c)}{Median heliocentric  distance in AU.}
   \tablenotetext{(d)}{Median geocentric  distance in AU.}
   \tablenotetext{(e)}{Median Solar phase angle (Sun--209P--Observer angle) in degree.}
   \tablenotetext{(f)}{Median position angle of normal vector with respect to the scattering plane in degree.}
   \tablenotetext{(g)}{Median position angle of the scattering plane in degree.}
   \tablenotetext{(h)}{Employed instruments.}
\label{tab:t1}
\end{deluxetable} 

%%
%% Table 2
\begin{deluxetable}{ccrrrrrrr} 
\tablecolumns{9} 
\tablewidth{0pc} 
\tablecaption{Degree of linear polarization and position angle of polarization for 209P} 
\tablehead{ 
%\colhead{}    &  \multicolumn{3}{c}{Non-shell Stars} &   \colhead{}   & 
%\multicolumn{2}{c}
%\cline{2-4} \cline{6-8} \\ 
\colhead{Date} & \colhead{Filter}  & \colhead{Ap$^{(a)}$} & \colhead{$P$$^{(b)}$}  & \colhead{$\sigma P$$^{(c)}$} & \colhead{$\theta_P^{(d)}$} 
& \colhead{$\sigma \theta_P^{(e)}$} & \colhead{$P_r$$^{(f)}$} &  \colhead{$\theta_r$$^{(g)}$}
}
\startdata
2014/04/23 & R$_\mathrm{C}$ & 3.03 & 30.7  & 3.0  &  3.9  & 2.8 & 30.6 & 2.6\\
2014/04/23 & R$_\mathrm{C}$ & 6.06 & 25.3  & 1.1  & 179.4 & 1.2 & 25.2 & 178.1\\
2014/04/23 & R$_\mathrm{C}$ & 9.09 & 27.8  & 4.8  &  2.2  & 4.9 & 27.8 & 0.9\\
2014/04/23 & R$_\mathrm{C}$ & 12.12 & 25.6  & 3.6  & 176.8 & 4.1 & 25.3 & 175.5\\
2014/04/23 & R$_\mathrm{C}$ & 15.15 & 21.2  & 7.4  & 172.4 & 52.0 & 20.2 & 171.1\\
2014/04/23 & J  & 2.59 & 33.8  & 2.2  & 1.5   & 1.8 & 33.8 & 0.2\\
2014/04/23 & J  & 5.18 & 31.7  & 5.5  & 0.7   & 5.0 & 31.7 & -0.6\\
2014/04/23 & J  & 7.77 & 30.5  & 5.5  & 176.4 & 5.1 & 30.1 & 175.1\\
2014/04/23 & J  & 10.36 & 26.3  & 3.3  & 172.9 & 3.5 & 25.2  & 171.6\\
2014/04/23 & J  & 12.95 & 20.6  & 4.1  & 1.3   & 5.7 & 20.6 & 0.0\\
2014/05/01 & R$_\mathrm{C}$ & 2.84 & 30.5  & 0.6  & 5.7   & 0.5 & 30.3 & 3.3\\
2014/05/01 & R$_\mathrm{C}$ & 7.06 & 32.1  & 3.4  & 1.5   & 3.1 & 32.1 & -0.9\\
2014/05/01 & R$_\mathrm{C}$ & 10.59 & 28.5  & 5.8  & 3.5   & 5.8 & 28.5 & 1.1\\
2014/05/01 & R$_\mathrm{C}$ & 14.12 & 36.2  & 6.4  & 179.3 & 5.0 & 36.0 & 176.9\\
2014/05/01 & R$_\mathrm{C}$ & 17.65 & 18.5  & 8.8  & 6.7   & 52.0 & 18.3 & 4.3\\
2014/05/01 & J  & 2.61 & 30.2  & 5.0  & 175.3 & 4.7 & 29.3 & 172.9\\
2014/05/01 & J  & 5.22 & 27.1  & 2.3  & 1.9   & 2.4 & 27.1 & -0.5\\
2014/05/01 & J  & 7.83 & 23.8  & 6.5  & 4.6   & 7.9 & 23.7 & 2.2\\
2014/05/01 & J  & 10.44 & 23.5  & 4.6  & 9.7   & 5.6 & 22.7 & 7.3\\
2014/05/01 & J  & 13.05 & 20.9  & 4.6  & 177.9 & 6.3 & 20.6 & 175.5\\
2014/05/04 & R$_\mathrm{C}$ & 3.53 & 30.12 & 0.18 &  2.01 & 0.21 & 30.10 & -1.16\\
2014/05/04 & R$_\mathrm{C}$ & 7.06 & 29.95 & 0.23 &  1.64 & 0.26 & 29.91 & -1.53\\
2014/05/04 & R$_\mathrm{C}$ & 10.59 & 29.62 & 0.24 &  1.41 & 0.27 & 29.56 & -1.76\\
2014/05/04 & R$_\mathrm{C}$ & 14.12 & 29.42 & 0.27 &  1.42 & 0.29 & 29.37 & -1.75\\
2014/05/04 & R$_\mathrm{C}$ & 17.65 & 29.86 & 0.35 &  1.40 & 0.36 & 29.80 & -1.77\\
2014/05/17 & R$_\mathrm{C}$ & 3.53 & 35.2  & 4.0  & 10.0  & 3.2 & 35.2 & -0.2\\
2014/05/17 & R$_\mathrm{C}$ & 7.06 & 37.3  & 6.8  & 10.8  & 5.2 & 37.3 & 0.6\\
2014/05/17 & R$_\mathrm{C}$ & 10.59 & 30.3  & 1.9  & 6.1   & 1.8 & 30.0 & -4.1\\
2014/05/17 & R$_\mathrm{C}$ & 14.12& 36.3  & 3.8  & 6.0   & 3.0 & 35.9 & -4.2\\
2014/05/17 & R$_\mathrm{C}$ & 17.65 & 32.1  & 3.5  & 5.1   & 3.1 & 31.6 & -5.1\\
2014/05/19 & R$_\mathrm{C}$ & 2.79 & 30.88 & 0.11 & 11.06 & 0.17 & 30.86 & -1.11\\
2014/05/19 & R$_\mathrm{C}$ & 5.58 & 30.47 & 0.12 & 11.03 & 0.17 & 30.45 & -1.14\\
2014/05/19 & R$_\mathrm{C}$ & 8.37 & 30.32 & 0.13 & 11.05 & 0.18 & 30.30 & -1.12\\
2014/05/19 & R$_\mathrm{C}$ & 11.16 & 30.35 & 0.14 & 10.99 & 0.18 & 30.32 & -1.18\\
2014/05/19 & R$_\mathrm{C}$ & 13.95 & 30.41 & 0.17 & 10.94 & 0.20 & 30.38 & -1.23\\
\enddata 
\tablenotetext{(a)}{Aperture radius in arcsec.}
\tablenotetext{(b)}{Polarization degree in percent.}
\tablenotetext{(c)}{Standard deviation of $P$ in percent.}
\tablenotetext{(d)}{Position angle of the strongest electric vector in degree.}
\tablenotetext{(e)}{Standard deviation of $\theta_P$ in degree.}
\tablenotetext{(f)}{Polarization degree referred to the scattering plane in percent.}
\tablenotetext{(g)}{Position angle referred to the scattering plane in degree.}

\label{tab:t2}
\end{deluxetable}

%%%%%%%%%%%%%%%%
%%   FIGURE 1  IMAGES %%
%%%%%%%%%%%%%%%%
\clearpage

\clearpage

%%%%%%%%%%%%%%%%%%%%%
%%   FIGURE 2 RADIAL PROFILES  %%
%%%%%%%%%%%%%%%%%%%%%

\clearpage

%%%%%%%%%%%%%%%%%%%%%%%%%%%%%%
%%   FIGURE 3 Polarization degree v.s. flux ratio  %%
%%%%%%%%%%%%%%%%%%%%%%%%%%%%%%

\clearpage 

%%%%%%%%%%%%%%%%%%%%%%%%%%%%%%%%%%%
%%   FIGURE 4 PHASE-ANGLE DEPENDENCE  + COMETS %%
%%%%%%%%%%%%%%%%%%%%%%%%%%%%%%%%%%%

\clearpage

%%%%%%%%%%%%%%%%%
%%   FIGURE 5 SPECTRUM %%
%%%%%%%%%%%%%%%%%

\clearpage

%%%%%%%%%%%%%%%%%%%%%%%%%%%%%%%%%%%
%%   FIGURE 6 PHASE-ANGLE DEPENDENCE  + ASTEROIDS %%
%%%%%%%%%%%%%%%%%%%%%%%%%%%%%%%%%%%

\clearpage

%%%%%%%%%%%%%%%%%%%%
%%   FIGURE 7 ALBEDO VS Pmax %%
%%%%%%%%%%%%%%%%%%%%

\begin{thebibliography}{}

\bibitem[Akitaya et al.(2014)]{Akitaya2014} Akitaya, H., Moritani, Y., Ui, T., et al.\ 2014, \procspie, 9147, 91474O 

\bibitem[Bagnulo et al.(2010)]{Bagnulo2010} Bagnulo S., Tozzi G.P., Boehnhardt H., et al.\ 2010, \aap,  514, 1

\bibitem[Belskaya et al.(2009)]{Belskaya2009} Belskaya I. N., Levasseur-Regourd A.-C., Cellino A. et al. \ 2009, \sovast, 199, 97

\bibitem[Belskaya et al.(1987)]{Belskaya1987} Belskaya I. N., Lupishko D. F., \& Shakhovskoj N. M.\ 1987, \sovast, 13, 219

\bibitem[Belskaya et al.(2005)]{Belskaya2005} Belskaya I. N., Shkuratov Yu. G., Efimov Yu. S. et al.\ 2005, \icarus, 178, 213

\bibitem[Bonev et al.(2008)]{Bonev2008} Bonev, T., Boehnhardt, H. \& Borisov, G. \ 2008, \aap, 480, 277

\bibitem[Brown(2014)]{Brown2014} Brown, P.\ 2014, Central Bureau Electronic Telegrams, 3886, 1

\bibitem[Brown et al.(1996)]{Brown1996} Brown, M.~E., Bouchez, A.~H., Spinrad, A.~H., \& Johns-Krull, C.~M.\ 1996, \aj, 112, 1197 

\bibitem[Canada-Assandri et al.(2012)]{Canada-Assandri2012} Canada-Assandri M., Gil-Hutton R., \& Benavidez P.\ 2012, 542, A11

\bibitem[Cellino et al.(1999)]{Cellino1999} Cellino, A., Hutton, R. Gil, Tedesco, E. F., et al.\ 1999, \icarus, 138, 129

\bibitem[Cellino et al.(2005)]{Cellino2005} Cellino, A., Yoshida, F., Anderlucci, E., et al.\ 2005, \icarus, 179, 297

\bibitem[Chernova et al.(1993)]{Chernova1993}  Chernova, G. P., Kiselev, N.N., \& Jockers, K. \ 1993, \icarus, 103, 133

\bibitem[Dollfus et al.(1989)]{Dollfus1989} Dollfus, A., Wolff, M., Geake, J.~E., Dougherty, L.~M., \& Lupishko, D.~F.\ 1989, Asteroids II, 594 

\bibitem[Dollfus(1998)]{Dollfus1998} Dollfus, A.\ 1998, \icarus, 136, 69 

\bibitem[Eaton et al.(1988)]{Eaton1988} Eaton, N., Scarrott, S. M., Warren-Smith, R. F.\ 1988, \icarus, 76, 270

\bibitem[Eaton et al.(1992)]{Eaton1992} Eaton, N., Scarrott, S. M., Gledhill, N. M.\ 1992, \mnras,  258, 384

\bibitem[Fern{\'a}ndez et al.(2000)]{Fernandez2000} Fern{\'a}ndez, Y.~R., Lisse, C.~M., Ulrich K{\"a}ufl, H., et al.\ 2000, \icarus, 147, 145 

\bibitem[Fornasier et al.(2015)]{Fornasier2015} Fornasier, S., Belskaya, I.~N., \& Perna, D.\ 2015, \icarus, 250, 280 

\bibitem[Fornasier et al.(2006)]{Fornasier2006} Fornasier S., Belskaya I., Shkuratov Yu.G., et al.,  2006, \icarus,  455, 371 

\bibitem[Furusho et al.(1999)]{Furusho1999} Furusho, R., Suzuki, B., Yamamoto, N., et al.\ 1999, \pasj, 51, 367 

\bibitem[Furusho et al.(2007)]{Furusho2007} Furusho, R., Ikeda, Y.,  Kinoshita, D., et al.\ 2007, \icarus, 190, 454 

\bibitem[Ganesh et al.(2009)]{Ganesh2009} Ganesh, S., Joshi, U. C., Baliyan, K. S.\ 2009, \icarus, 201, 666

\bibitem[Gil-Hutton \& Canada-Assandri(2012)]{Gil-Hutton2012}  Gil-Hutton, R. \& Canada-Assandri, M.\ 2012, \aap, 539, A115

\bibitem[Gil-Hutton et al.(2014)]{GilHutton2014} Gil-Hutton, R., Cellino, A., \& Bendjoya, P.\ 2014, \aap, 569, AA122 

\bibitem[Goidet-Devel et al.(1995)]{Goidet-Devel1995} Goidet-Devel, B., Renard, J.~B., \& Levasseur-Regourd, A.~C.\ 1995, \planss, 43, 779

\bibitem[Hadamcik \& Levasseur-Regourd(2003a)]{Hadamcik2003a} Hadamcik, E., \& Levasseur-Regourd, A. ~C.\ 2003, \jqsrt, 79-80, 679

\bibitem[Hadamcik \& Levasseur-Regourd(2003b)]{Hadamcik2003b} Hadamcik, E., \& Levasseur-Regourd, A.~C.\ 2003, \aap, 403, 757 

\bibitem[Hadamcik \& Levasseur-Regourd(2003c)]{Hadamcik2003c} Hadamcik, E., \& Levasseur-Regourd, A.~C.\ 2003, \icarus, 166, 188

\bibitem[Hadamcik \& Levasseur-Regourd(2009)]{Hadamcik2009} Hadamcik, E., \& Levasseur-Regourd, A.~C.\ 2009, \planss, 57, 1118

\bibitem[Hadamcik et al.(1997)]{Hadamcik1997} Hadamcik, E., Levassuer-Regourd, A. C., \& Renard, J. B. \ 1997 , Earth, Moon, and Planets, 78, 365

\bibitem[Hadamcik et al.(2006)]{Hadamcik2006} Hadamcik, E., Renard, J. B., Levasseur-Regourd, A. C. et al., \ 2006,  \jqsrt, 100, 143

\bibitem[Hadamcik et al.(2002)]{Hadamcik2002} Hadamcik, E., Renard, J. B., Worms, J. C., et al., \ 2002, \icarus, 155, 497

\bibitem[Hadamcik et al.(2013)]{Hadamcik2013} Hadamcik, E., Sen, A.~K., Levasseur-Regourd, A.~C., et al.\ 2013, \icarus, 222, 774

\bibitem[Hanner(2003)]{Hanner2003} Hanner, M., \ 2003, \jqsrt, 79-80, 695

\bibitem[Harris et al.(2007)]{Harris2007} Harris, A. W., Mueller, M., Delb\'{o}, M., et al.\ 2007, \icarus, 188, 414

\bibitem[Hergenrother(2014)]{Hergenrother2014} Hergenrother, C.\ 2014, Central Bureau Electronic Telegrams, 3870, 1 

\bibitem[Howell et al.(2014)]{Howell2014} Howell, E.~S., Nolan, M.~C., Taylor, P.~A., et al.\ 2014, AAS/Division for Planetary Sciences Meeting Abstracts, 46, \#209.24 

\bibitem[Ishiguro et al.(2015)]{Ishiguro2015} Ishiguro, M., Kuroda,  D., Hanayama, H., et al.\ 2015, \apjl, 798, L34 

\bibitem[Ishiguro et al.(1997)]{Ishiguro1997} Ishiguro, M., Nakayama, H., Kogachi, M., et al.\ 1997, \pasj, 49, L31 

\bibitem[Jenniskens \& Lyytinen(2014)]{Jenniskens2014} Jenniskens, P., \& Lyytinen, E.\ 2014, Central Bureau Electronic Telegrams, 3869, 1 

\bibitem[Jewitt(2004)]{Jewitt2004} Jewitt, D.\ 2004, \aj, 128,  3061

\bibitem[Jorkers et al.(2005)]{Jorkers2005} Jockers, K., Kiselev, N., Bonev, T., et al. \aap, 441, 773

\bibitem[Jones \& Gehrz(2000)]{Jones2000} Jones, T.~J., \& Gehrz, R.~D.\ 2000, \icarus, 143, 338 

\bibitem[Joshi et al.(2003)]{Joshi2003} Joshi, U. C., Baliyan, K. S., Ganesh, S.\ 2003, \aap, 405, 1129

\bibitem[Kraemer, et al.(2005)]{Kraemer2005} Kraemer, K. E., Lisse, C. M., Price, Stephan D.,  et al.\ 2005, \aj, 130, 2263 

\bibitem[Kawabata et al.(1999)]{Kawabata1999} Kawabata, K. S., Okazaki, A., Akitaya., H., et al.\ 1999, \pasp, 111, 898 

\bibitem[Kikuchi(2006)]{Kikuchi2006} Kikuchi, S.\ 2006, \jqsrt, 100, 179

\bibitem[Kikuchi et al.(1987)]{Kikuchi1987} Kikuchi, S., Mikami, Y., Mukai, T., et al.\ 1987, \aap, 187, 689

\bibitem[Kikuchi et al.(1989)]{Kikuchi1989} Kikuchi, S., Mikami, Y., Mukai, T., et al.\ 1989, \aap, 214, 386

\bibitem[Kim et al.(2014)]{Kim2014} Kim, Y., Ishiguro, M., Usui, F.\ 2014, \apj, 789, 151

\bibitem[Kiselev \& Chernova(1978)]{Kiselev1978} Kiselev, N. N., Chernova, G. P.\ 1978, Soviet Astronomy Letters, 55, 1064

\bibitem[Kiselev et al.(2004)]{Kiselev2004} Kiselev, N.~N., Jockers, K., \& Bonevc, T.\ 2001, Solar System Research, 35, 480

\bibitem[Kiselev et al.(2001)]{Kiselev2001} Kiselev, N.~N., Jockers, K., Rosenbush, V.~K., \& Korsun, P.~P.\ 2001, Solar System Research, 35, 480 

\bibitem[Kiselev et al.(1990)]{Kiselev1990} Kiselev, N. N., Lupishko, D. F., Chernova, G. P., et al.\ 1990, Kinematika i Fizika Nebesnykh Tel, 6, 77

\bibitem[Kiselev et al.(1999)]{Kiselev1999} Kiselev, N.~N., Rosenbush, V.~K., \& Jockers, K.\ 1999, \icarus, 140, 464 

\bibitem[Kiselev et al.(2002)]{Kiselev2002} Kiselev, N. N., Rosenbush, V. K., Jockers K., et al.\ 2002, in: Proceedings of Asteroids, Comets, Meteors - ACM 2002 ed. B. Warmbein (Berlin, Germany : ESA SP-500), 887

\bibitem[Kiselev et al.(2005)]{Kiselev2005} Kiselev, N., Velichko, S., Jockers, K., et al.\ 2005, Earth, Moon, and Planets, 97, 365

\bibitem[Kolokolova et al.(2007)]{Kolokolova2007} Kolokolova, L., Kimura, H., Kiselev, N., \& Rosenbush, V.\ 2007, \aap, 463, 1189 

\bibitem[Lamy et al.(2004)]{Lamy2004} Lamy, P. L., Toth, I., Fernandez, Y. R., \& Weaver, H. A.\  2004, Comets II, 223

\bibitem[Levasseur-Regourd et al.(1996)]{Levasseur1996} Levasseur-Regourd, A.~C., Hadamcik, E., \& Renard, J.~B.\ 1996, \aap, 313, 327 

\bibitem[Lumme \& Muinonen(1993)]{Lumme1993} Lumme, K., \& Muinonen, K.~O.\ 1993, LPI Contributions, 810, 194 

\bibitem[Lupishko(2014)]{Lupishko2014} Lupishko, D.\ 2014, NASA Planetary Data System, 215,  

\bibitem[Lupishko et al.(1994)]{Lupishko1994} Lupishko D. F., Kiselev N. N., Chernova G. P.,  et al.\ 1994, Kinematika Fiz. Nebesn. Tel., 10, 40-44

\bibitem[Masiero et al.(2012)]{Masiero2012} Masiero, J. R., Mainzer, A. K., Grav, T., et al.\  2012, \apj, 749, 104

\bibitem[Michalsky(1981)]{Michalsky1981} Michalsky J. J.\ 1981,  \icarus, 47, 388

\bibitem[Moffat (1969)]{Moffat1969} Moffat, A. F. J.\ 1969, \aap, 3, 455 

\bibitem[Mukai et al.(1997)]{Mukai1997} Mukai, T., Iwata, T.,  Kikuchi, S., et al.\ 1997, \icarus, 127, 452 

\bibitem[Myers(1985)]{Myers1985} Myers, R. V.\ 1985, \icarus, 63, 206

\bibitem[Nakayama et al.(2000)]{Nakayama2000} Nakayama H., Fujii Y., Ishiguro M., et al.\ 2000, \icarus, 146,  220-231

\bibitem[Noland et al.(1973)]{Noland1973}  Noland, M., Veverka, J., \& Pollack, J. B.\ 1973, \icarus, 20, 490

\bibitem[Penttil\"{a} et al.(2005)]{Penttila2005} Penttil\"{a}, A., Lumme, K., Hadamcik, E., et al.\ 2005, \aap, 432, 1081

\bibitem[Rosenbush et al.(1994)]{Rosenbush1994} Rosenbush, V. K., Rosenbush A. E., Dement'ev M. S.\ 1994,  \icarus, 108, 81

\bibitem[Schleicher(2014)]{Schleicher2014} Schleicher, D.\ 2014, Central Bureau Electronic Telegrams, 3880, 1 

\bibitem[Shkuratov et al.(2002)]{Shkuratov2002} Shkuratov, Y., Ovcharenko, A., Zubko, E., et al.\ 2002, \icarus, 159, 396 

\bibitem[Shevchenko \& Skobeleva(1995)]{Shevchenko1995} Shevchenko, V. V., \& Skobeleva, T. P.\ 1995, Solar System Research, 29, 74 

\bibitem[Tinbergen(1996)]{Tinbergen1996} Tinbergen, J. \ 1996,  Astronomical Polarimetry (Cambridge University Press), iSBN: 0521475317

\bibitem[Thomas et al.(1996)]{Thomas1996} Thomas, P. C., Adinolfi, D., Helfenstein, P., et al. \ 1996, \icarus, 123, 536

\bibitem[Tozzi et al.(1997)]{Tozzi1997} Tozzi, G. P.,  Cimatti, A., di Serego Alighieri, S., et al. \ 1997, \planss, 45, 535

\bibitem[Watanabe et al.(2012)]{Watanabe2012} Watanabe, M., Takahashi, Y., Sato, M., et al.\ 2012, \procspie, 8446, 84462O

\bibitem[Zellner \& Capen(1974)]{Zellner1974a} Zellner, B., \& Capen, R. C. \ 1974, \icarus, 23, 437

\bibitem[Zellner et al.(1974)]{Zellner1974b} Gehrels T., \& Gradie J. \ 1974, \aj, 79, 1100

\bibitem[Zellner \& Gradie(1976)]{Zellner1976} Zellner, B.,  \& Gradie J. \ 1976, \aj, 81, 262

\bibitem[Zubko et al.(2011)]{Zubko2011} Zubko, E., Furusho, R., Kawabata, K., et al. \ 2011, \jqsrt, 112, 1848 



\end{thebibliography}
\end{document}